\documentclass[12pt]{article}
\include{epsfig}
\include{graphicx}

\begin{document}

\title{Transient situations in traffic flow:
Modelling the Mexico City Cuernavaca Highway}

\author{J.A. del R\'{\i}o \\
Centro de Investigaci\'on en Energ\'{\i}a \\
Universidad Nacional Aut\'onoma de M\'exico\\
A.P. 34, 62580 Temixco, Mor. M\'exico
\\
M.E. L\'arraga\\
Facultad de Ciencias \\
Universidad Aut\'onoma del Estado de Morelos\\
Cuernavaca, Mor. M\'exico}

\maketitle

\begin{abstract}
In this paper a recent variable anticipation cellular automata
model for single-lane traffic flow
is extended to analyze the situation of free and congested flow in
the Highway from Mexico City to Cuernavaca. This highway presents
free flow in standard days; but in the returning day of long weekends
or holidays it exhibits
congested flow and in rush hours jamming appears. We illustrate
how our CA model for traffic flow can deal appropriately  with
transient situations and can be used to search new alternatives
that allow to improve the traffic flow in Mexican highways.

\end{abstract}


\section{Introduction}

Nowadays the life style of people living in big cities implies the
requirement of fast individual or collective mobility. For this
reason, the traffic flow has important relevance in the economy of
a country.  This has generated a constant search for models that
capture the essentials of the dynamics of the transportation
system. Clearly, traffic flow is a system far from equilibrium
where a description of the transitions between jammed, congested,
synchronized (platoons) and free flow is required \cite{chow}.
Recently progress has been made in understanding traffic
flow using statistical physics and cellular automata models
\cite{chow}, \cite{nag1}, \cite{Hel}. These models have turned out as
excellent tools
for the simulation of large scale traffic networks.

Cellular automata (CA) are agent-based simulations for complex natural
systems containing large numbers of simple identical components
which interact locally. There is an orderly lattice of sites, each
with a finite set of possible values, in the cellular automata
model. The value of the sites evolves synchronously in discrete
time steps according to a particular set of rules. The value of a
particular site is determined by the previous values of a
neighbourhood of sites around it. This really simplifies the study
of a complex system with many components and eases the process of
analysis. Without the CA model, there would be infinitely many
possible connections between components. The Nagel-Schreckenberg
(NaSch) model \cite{nag1} considered the effects of acceleration and delay of
vehicles with high speed. This model captures the realistic
traffic situations where the car accelerates and decelerates.
However, in this model, a simple single-lane CA is set up and 
anticipation effects that play a important role
to elucidate the behavior of real traffic flow are not considered.

Different models have been performed modifying the NaSch model.
Recently, a new CA model
including variable anticipation showed the existence of platoons
travelling forwards depending on the anticipation strength, therefore
opening new aspects to analyze in traffic flow \cite{larraga03}.

While a single-lane model gives a good analysis of traffic flow, it is
not sufficient to depict the real world situations. As a result,
two-lane traffic models based on cellular automata have been
proposed in order to incorporate the rules able to
deal systematically with multi-lane highways \cite{Hel,Kno1,nagel98}.
However the
role of anticipation and multi-lane factors need to be deeply
described. One can describe the two-lane flow using two
single-lane systems and defining the corresponding  lane changing
rules for coupled sinks and sources for each lane \cite{chow}.

Although most studies on CA models for traffic flow have been done
for systems with periodic boundary conditions, open boundaries are
relevant for many realistic situations in traffic where the number
of vehicles can change, e.g., due to ramps or toll booths, as in the
Mexico City-Cuernavaca Highway (MCH).

In this paper, firstly we extend a variable anticipation model
proposed in ref. \cite{larraga03} to consider multi-lane highways. The
aim is to analyze the behavior of the MCH traffic
flow to propose new alternatives to improve its management of
traffic flow, where a congested flow is observed on days-off. For
this purpose, we consider that vehicular traffic on a highway is
controlled by a mixture of bulk and boundary effects caused by on-
and off- ramps, varying number of lanes, speed limits and types of
vehicles. The inclusion of this variety also opens new transitions
and different behavior.

The organization of the paper is the following: In next section we
present the modification of the one-lane variable anticipation
model for traffic flow \cite{larraga03} to describe multi-lane
traffic flow under asymmetric lane change rules. Then we describe
the schematic of the MCH considering
entrances, exits and topology. We the proceed to present the main
results of our simulations. Finally we close the paper with some
concluding remarks.


\section{The model}
\label{modelo}

In this section, we briefly describe a recently introduced CA
traffic flow model with variable anticipation \cite{larraga03},
where a single-lane CA is set up. This model
has shown that it is able to reproduce the most important
phenomena observed in real traffic flow. After defining the
single-lane model, a very simple extension to a model multi-lane
traffic flow is presented.

\subsection{Single lane model}

For the single-lane model, the length of the highway is divided in
uniforms cells. Each cell represents the space that is enough to
accommodate one vehicle. In the simulation model of this paper, the
length of each cell is $7.5m$. Each cell can be empty, or occupied
by a vehicle $i$ of $N$ existing in a network, with an integer
velocity $v_i \in \{0,1, \cdots, v_{max}\}$. The maximum velocity
of vehicle $v_{max}$ is defined to be 5 cell-lengths, which
corresponds to $37.5cell/s$

The number of unoccupied cells between a vehicle $i$ and its
preceding vehicle is defined as ''gap'' (denoted $d_i$). In a
single-lane CA model for traffic flow, the following rules are
applied to all $N$ vehicles on the lattice in each iteration,
which corresponds to 1 second of real time:

\begin{description}
\item[R1:] Acceleration \\
If $v_{i}<v_{max}$, the velocity of the car $i$ is increased by one, i.e.,
$$
v_{i}\rightarrow \min(v_{i}+1,v_{max}).
$$
\item[R2:] Randomization \\
If $v_{i}>0$, the velocity of car $i$
\ is decreased randomly by one unit with probability $R$, i.e.,
$$
v_{i}\rightarrow \max(v_{i}-1,0) \quad {\rm with\ probability}~R.
$$
\item[R3:] Deceleration \\
If $d_{i}^{s}<v_{i}$, where
$$
d_{i}^{s}=d_{i}+\left[(1-\alpha )\cdot v_{p}+\frac{1}{2}\right],
$$
with a parameter $0\leq \alpha \leq 1$, the velocity of
car $i$\ is reduced to $d_{i}^{s}$. $[x]$ denotes the integer
part of $x$, i.e.\ $[x+\frac{1}{2}]$ corresponds to rounding $x$ to
the next integer value.\\
The new velocity of the vehicle $i$ is therefore
$$
v_{i}\rightarrow \min(v_{i},d_{i}^{s}).
$$
\item[R4:] Vehicle movement \\
Each car is moving forward according
to its new velocity determined in steps 1-3, i.e.,
$$
x_{i}\rightarrow x_{i}+v_{i}.
$$
\end{description}
Rules $R1$, $R2$ and $R3$ are designed to update the velocity of vehicles;
rule $R4$ updates position.

While the single-lane model gives a good analysis of traffic flow, it
is not sufficient to depict the real-world situations for highways
with more than one lane. As a result, to carry out the aim of this paper
a multi-lane model should be defined. In order to extend the model
to multi-lane traffic one has to introduce lane-changing rules.

\subsection{Multi-lane model}

Therefore, in addition to the rules that apply to the single-lane
model, there is also a rule for ''lane-changing'', which is a
special characteristic of multi-lane traffic flow. In principle,
all lane changing rule sets of cellular automaton models for
traffic flow are formulated analogously. First, a vehicle needs an
incentive to change a lane. Second, a lane change is only possible
if some safety constraints are fulfilled. In this way, we have to
first consider security such that there must be enough space on
the target lane. Technically, one can say that the gap on the
target lane in front of (behind) the vehicle that wants to change
lanes should be safe. This is to ensure that after it has changed
lane, neither will it crash into the vehicle in front of it, nor
be crashed into by the following vehicles, in the target lane.

Besides, asymmetry for lane-changing is introduced if one applies
different criteria for the change from left to right and right to
left due to legal constraints. For instance, in Mexico, the right
lane has to be used by default and passing has to be on the left;
whereas in the United States, passing can be both on left or
right.

A system update is performed in two sub-steps: In the first step
the cars change the lanes according to the lane changing rules and
do not move. In the second step, the cars move according to the
calculated velocity. Both sub-steps are performed in parallel for
all vehicles.

In this paper, we define the lane changing strategy based on
driving laws for Mexico highways that lead to an asymmetry between
the lanes:
\begin{description}
\item[(i)] The right lane preference is enforced by the legal
regulation to use the right lane as often as possible.

\item[(ii)] The right lane overtaking ban prohibits a car driving
on the left lane to use the right lane to overtake a car which is driving
on the left lane.
\end{description}

In order to reproduce the multi-lane traffic flow, we introduce
the right lane preference and the right lane overtaking ban
simultaneously. Vehicles are still allowed to overtake their
predecessor on the left lane but the left lane should be
preferred. The lane change rules are then as follows:
\begin{description}
\item \textbf{Right lane to left lane}

\begin{description}

 \item [Incentive Criterion] A
vehicle will try to change from the right to the left lane only if
the interaction with its predecessor on its lane implies that it
will brake for safety or just for comfort reasons. It is expressed
by: $$v_{i}>d_{i}^{s},$$
where $v_i$ is the velocity of the vehicle that is changing of lane.
$d_{i}^{s}=d_{i}+\left[(1-\alpha )\cdot v_{p}+\frac{1}{2}\right]$,
where $v_p$ is the velocity of the preceding vehicle it on its
actual lane.

\item[Safety Criterion] This change will be possible if the gap
between successor and predecessor on the target lane is sufficient
and safe. The movement of the preceding vehicles  both the target
lane and at the original lane is anticipated (based on safety distance
$d^s$). If
$$\left(d_{pr}^{s}\geq v_{i}\right) \quad and \quad  \left(d_{su}^{s}\geq v_{su}\right),$$
with $d_{su}$ and $v_{su}$ being the gap to the succeeding vehicle on the
target lane and its velocity, respectively. $d_{pr}$ denotes de
gap and  $d_{pr}^{s}=d_{pr}+\left[(1-\alpha )\cdot v_{p}^{\prime}
+\frac{1}{2}\right]$ denotes the safety gap to the preceding
vehicle on the destination lane (where $v_{pr}$ is the velocity of
the predecessor on the destination lane).

\end{description}


\item \textbf{Left lane to right lane}


\begin{description}
\item[Incentive criterion] We reduced the ability to change from
the left to the right lane so that vehicles on the left lane
change back to the right lane only if there is a sufficient gap on
both, the right and the left lane:
$$\left(d_{pr}^s/v_i \geq 3.0\right) \quad and \quad \left(\left( d_i^s/v_i
\geq 3 \right) \quad or \quad
\left(v_{i}>d_{i}^{s}\right)\right),$$
Here we have
introduced two times $d_{i}^{s}/v_{i}$ and $d_{i}^{s\prime
}/v_{i}$ that give the time intervals needed for a given vehicle to reach the position
of its predecessor and for its predecessor on the destination lane.
Since defined times take the velocity of the vehicles into
account, slow vehicles are allowed to change the lane even at
small distances.

\item[Safety criterion] Safety distance to the succeeding vehicle
in the target lane is larger that the velocity of the succeeding
vehicle to avoid  it braking and producing an accident.
$$\left(d_{su}^{s}\geq v_{su}\right)$$

\end{description}

\end{description}

With these rules, we may describe the behavior of the multi-lane model. In the
next section, this multi-lane model will be applied to analyze the
Mexico City-Cuernavaca Highway (MCH) traffic flow.  MCH is one of
the most used highway on days-off. Its complex topology and high
flow produces congestions and a great increase of travel time.

\section{Modelling the Mexico City-Cuernavaca Highway}
\label{carretera}

MCH has a length of 64 Km and is composed by two flow directions,
so called DI1 and DI2. However in this work we only will present
simulation results corresponding to the direction D2 (from
Cuernavaca to Mexico City) that is one of the most interesting to
analyze.

The vehicular traffic on the MCH is controlled by boundary effects
caused by on- and off- ramps and toll booths. Thus, simulating
the MCH we will consider the multi-lane model defined in this work
on a lattice with open boundary conditions. Throughout the paper
we assume that the vehicles move from left to the right.
Therefore, vehicle injection is done at the left boundary
(corresponding to in-flow into the road segment) and removal of
vehicles at the right boundary (corresponding to outflow)
the behavior  of which is determined by a tool booth. We pay special
attention to simulate adequately this behavior.

\begin{figure}[htbp]
  \includegraphics[scale=0.50]{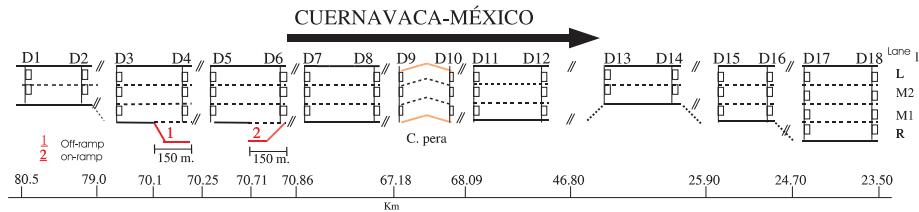}
  \caption{Schematic representation of Mexico City-Cuernavaca
highway, direction DI2}
  \label{cue_mex}
\end{figure}

\begin{figure}[htbp]
  \includegraphics[scale=0.80]{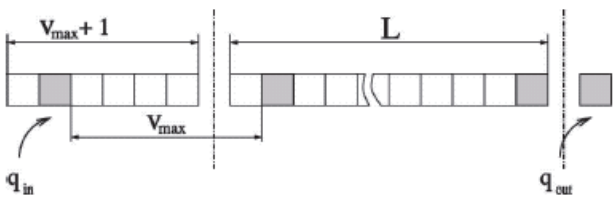}
  \caption{Schematic representation of the analyzed system.
  vehicles move from left to right, and are represented by dark cells,
whereas empty cells are white. The left boundary is given by a
mini-system consisting of $v_{max}$ cells. This reservoir is
occupied by at most one car with probability $q_{in}$. The right
boundary consists of a single cell occupied with probability
$q_{out}$.} \label{front1}
\end{figure}

A schematic representation of the analyzed system boundaries is
depicted in Fig. \ref{front1} \footnote{This representation is
based on the definition done in \cite{andreas}}. We expanded the
width of the left boundary from one single cell to a mini-system
of width $v_{max}+1$. The allocation of the
mini-system (left boundary) has to be updated every time step
before the vehicles of the complete system. The update procedure
consists of two steps. If one cell of the mini-system is occupied,
it has to be emptied first. Then a vehicle with initial velocity
$v_{max}$ (established according to the type of vehicle) is inserted
with probability $q_{in}$. Its position has to satisfy the
following conditions: (i) The distance to the first car in the
main system is at least equal to its corresponding maximum
velocity $v_{max}$, and (ii) the distance to the main system has
to be minimal, i.e., if no vehicle is present in the main system
within the first $v_{max}$ cells, the first cell of the boundary
is occupied.

On the other hand, the right boundary is realized by a single cell
linked to the end of the system. Here the update is applied
similar to the case of the left boundary before the general
vehicle update procedure. First, the right boundary is cleared (if
necessary) and then occupied with probability $q_{out}$. This
corresponds to an probability of $1-q_{out}$ that is established
based on the maximum outflow (by second and lane) allowed by the
toll booth. At last, cars are removed if their velocity is large
enough to reach at least the (empty) boundary cell. According to
reports of the Mexican Highway Office, in this analysis we take
$q_{out}=0.167$ (outflow is approximately $3000veh/hr$).

The on- and off-ramps existing in the MCH are
implemented as connected parts of the lattice where the vehicles
may enter or leave the system. The ramps activity is characterized
by the number of vehicles that enter (or leave) the main system
$q_{inr}$ ($q_{our}$) by time unit. Each time-step of simulation
can add (or remove) vehicles as a function of the probability
$q_{inr}$. The length chosen for the on- and off-ramps and the
distance between them is motivated by the dimensions found on the actual
MCH: $L_{ramp} = 20$ is the length of the ramps in units of the
lattice constant (identified with 7.5 meters). The first cell
corresponding to the on-ramp (out-ramp) is placed in $x_{on}$
($x_{off}$). for this reason in this paper we only analyze a section of
highway (not a complete network), we made a successive search into
the region corresponding to the on-ramp (from $x_{on}$ to
$x_{on}+L_{ramp}$) up to where empty car cell is found. Then a vehicle will
be inserted in that cell. Note that this condition for insertion
is possible in CA models because the defined deceleration process
allows it. Out-ramp works in a similar way. In the simulation
presented here, we consider 20\% of $q_{in}$ as inflow in on-ramp
and 5\% of $q_{in}$ as outflow in the out-ramp.

Moreover, as we can see from the scheme for the MCH shown in Figure
\ref{cue_mex}, this presents widening and narrowing points and
also there is a dangerous curve where vehicles are forced to
slow their velocities, thus in this section $v_{max}$ was reduced
for both slow and fast vehicles. In this scheme 
the places referenced by D1,D2,...D18 that we will use to measure
the fundamental variables of traffic flow in simulations are also
indicated. It is
important to say that these places are virtual, that is, these do
not exist in the reality and they were defined based on the MCH
topology (number of lanes, curves).

\section{Simulation results}
\label{results}

In order to simulate the MCH according to the model and
descriptions done in previous sections, we consider two types of
vehicles as a function of the value of the maximum velocity
$v_{vmax}$, slow and fast vehicles. The maximum speed of slow and
fast vehicles is $v_{max}=3$ and $v_{max}=5$, respectively. We
keep the cell size and time-step as those defined for the single
lane model, that is $7.5m$ and $1s$ respectively. In this way, the
number of cells of the system is obtained based on the real length
of the MCH:
$$
L= \textnormal{length of the road (m) /cell length}
$$
Besides, due to the agreement of data of simulation with the 
empirical fundamental diagram for single-lane model
\cite{larraga03} for not-automated vehicles, the values for
parameters $R$ and $\alpha$ are $R=0.2$ and $\alpha=0.75$.

On the other hand, experience tells us that the traffic flow in
the MCH is almost free flow in standard days, but during days-off
or holidays jamming is observed in the direction DI2. This
observed behavior is due to the fact that the toll can deal with
3000 veh/hr as a maximum. For this reason, we simulate the
afternoon of a last day on a long weekend or holidays as it is
described in Table 1. The aim is to illustrate how our CA model for
traffic flow can deal appropriately  with transient situations and
can be used to search new alternatives that allow to improve the
flow on the MCH.

\begin{table}
\begin{tabular}{|c|c|c|}\hline
\textbf{Period} & \textbf{Time Interval} & Inflow (left boundary) \\
\hline  \hline 1 & 60 mins & 2000 \textit{veh/hr} \\
\hline 2 & 15 mins & 2500 \textit{veh/hr} \\
\hline 3 & 15 mins & 3000 \textit{veh/hr} \\
\hline 4 & 60 mins & 3500 \textit{veh/hr} \\
\hline 5 & 60 mins & 3000 \textit{veh/hr} \\
\hline 6 & 30 mins & 2500 \textit{veh/hr} \\
\hline 7 & 110 mins & 2000 \textit{veh/hr} \\ \hline
\end{tabular}
\caption{Temporal flow variations in our simulation. Time interval
is the time (in minutes) that the inflow is kept constant}
\label{t1}
\end{table}

In this way, simulation starts with an inflow probability
$q_{in}=0.55$ (2000 veh/hr/lanes). This inflow stays by $10*L$
time-steps until a steady state is reached. After this steady
state, inflow is varied according to the time periods shown in
Table\ref{t1}. Note that the total time resulting from adding periods
is equal to 6 hrs. that corresponding  to an afternoon of a returning
day of holidays (after 16:00 hr) where rush hours are reached.

\begin{figure}[htbp]
  \includegraphics[height=.5\textheight]{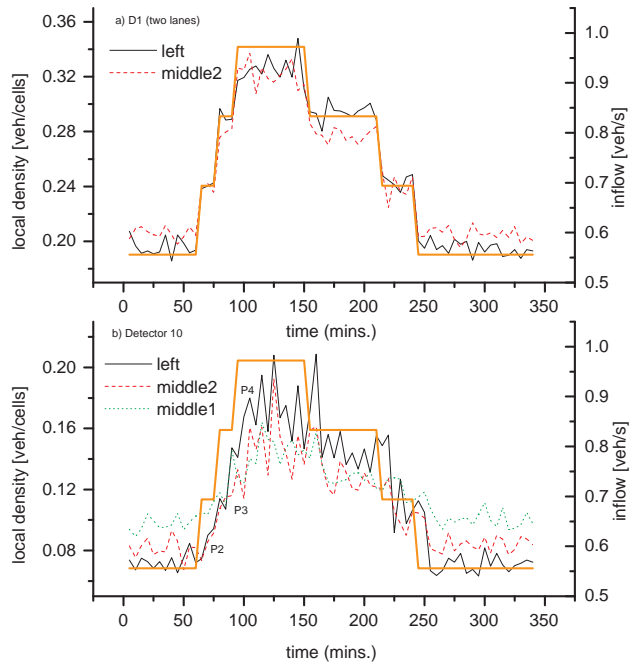}
  \caption{Local density averaged each five minutes respect to time. Time on
horizontal axis corresponds to the minutes passed after the first
period of Table \ref{t1} started. The right axis indicates the
variations of inflow. The thick solid line indicates inflow in D1 and
thin lines correspond to local densities measured in D1 (a) and
in D10 (b)} \label{tran12ab}
\end{figure}

In figure \ref{tran12ab}a) we show the local averaged density as a
function of time (left axis) measured in detector D2. Time on the
horizontal axis corresponds to the minutes passed from the first
period of Table \ref{t1} started. The right axis indicates the
variations of inflow. Note that the density varies according to
the inflow. Of course, these inflow variations are noted by
successive sections some time later, as we can see from figure
\ref{tran12ab}b) corresponding to the detector D10. Results
indicate that the D10 lasts approximately 15 minutes to
note an increase of the inflow. Besides, it is very important to
note that for both sections (D1 and D10) the jamming never is
reached under the inflow proposed in Table \ref{t1}.

\begin{figure}[htbp]
  \includegraphics[height=.5\textheight]{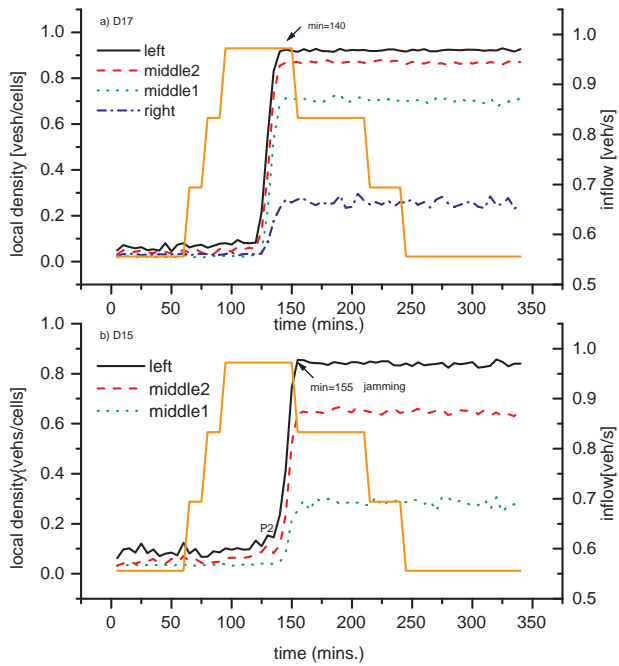}
  \caption{Local density averaged each five minutes respect to time. Time on
horizontal axis corresponds to the minutes passed after the first
period of Table \ref{t1} started. The right axis indicates the
variations of inflow. The thick solid line indicates inflow in D1 and
thin lines correspond to  local densities measured in D17 (before
toll booth) (a) and in D15 (b)} \label{tran34ab}
\end{figure}

On the other hand, in figure \ref{tran34ab}a), we show the local
average density in the detector D17. This position is just placed
before the toll booth. We observe clearly that before minute 135
the state of flow is free (low density), but five minutes later
the jamming appears. When the jamming appears it propagates
upstream of the detector and this produces in turn backward jamming in
the preceding sections. We can observe this situation in detector
D15 (see figure \ref{tran34ab}a)), where a jamming is observed in
minute 155, that is, 15 minutes later than the jamming that occurred
in D17. The topology of the MCH and a bad toll system do not allow
that the free flow will be recovered, so that the highway presents a
jamming and the travel time is increased up to 100$\%$ density.

It is important to emphasize that all the parameters used in this
simulation are based on the available information about the MCH in
Mexico. There are not empirical data to analyze
not working days in a microscopic form. We think that inanalysis similar to
the one presented in this paper could be used for planning, designing,
and operating of some Mexican highways where  a good
service level is reported, but the experience shows jamming in days-off.

Finally, simulations adding extra lanes indicate that
these additions do not solve the problem of traffic jams during the
returning days. However, we have done simulations of the MCH without considering the
toll booth and the results obtained indicated a free-flow state
whenever there does not exist another incident.
The same conclusion applies to the analysis of the opposite direction
(Mexico City to Cuernavaca) free flow regime is always present if there
is not an accident. An important point to be stressed is that with this
multilane model it is possible
to take into account more
complex situations (as accidents), with some modifications in specific
points on the highway topology. Also it is possible to analyze the impact of
strategies of safety.
For example, the selection of points on the highway where emergency stations may
be placed minimizing
their responding time.

\section{Concluding Remarks}
\label{conclu}

In this paper, a multi-lane extension of a recent CA model \cite{larraga03}
with
variable anticipation for single-lane traffic flow is presented.
The analysis focuses on the description of the traffic flow on
the Mexico-City Cuernavaca highway. Simulation results show that
the congestion observed on days-off in this highway is not only a
consequence of its complex topology. The main problem lies in the fact
that there is an inefficient system of toll. Results indicate that
the application of the CA model presented here could be used to
investigate new alternatives to improve the management of traffic
flow in Mexican Highways, where there does not exist this kind of
analysis.

Traffic simulations have shown to be a fascinating source of new kind of
phase transitions \cite{chow} \cite{larraga03} in complex systems.
Moreover, in this paper we have also illustrated how this kind of
simulations can be useful to test new strategies in solving actual
traffic problems.


\textbf{Acknowledgements}
We thank Prof. Leopodo Garc\'{\i}a Col\'{\i}n for suggesting the analysis
of the traffic problem of the Mexico City Cuernavaca Highway.


\bibliographystyle{aipproc}   

\bibliography{traffic}

\IfFileExists{\jobname.bbl}{}
 {\typeout{}
  \typeout{******************************************}
  \typeout{** Please run "bibtex \jobname" to optain}
  \typeout{** the bibliography and then re-run LaTeX}
  \typeout{** twice to fix the references!}
  \typeout{******************************************}
  \typeout{}
 }

\end{document}